\documentclass[aps,preprint,epsf,epsfig,amsmath]{revtex4}
\usepackage{graphicx,pdflscape,epsf,epsfig,amsmath}
\usepackage{alltt,dsfont,amsmath,amssymb,bm}
\usepackage[dvipsnames]{xcolor}
\usepackage{bm}
\usepackage{bbold}
\usepackage{mathtools}
\usepackage{alltt,dsfont}
\usepackage{appendix}
\usepackage{amsmath,amssymb}
\usepackage{hyperref}
\hypersetup{
    colorlinks,
    citecolor=red,
    filecolor=blue,
    linkcolor=blue,
    urlcolor=blue
}

\usepackage{atveryend}
\makeatletter
\let\origcitation\citation
\AtEndDocument{\def\mycites{}%
  \def\citation#1{\g@addto@macro\mycites{#1^^J}\origcitation{#1}}}
\AtVeryEndDocument{\newwrite\citeout\immediate\openout\citeout=\jobname.cit
  \immediate\write\citeout{\mycites}\immediate\closeout\citeout}
\makeatother

\newcommand{\iden}{ \mathds{ 1}}

\newcommand{\Q}{{\cal Q}}

\newcommand{\nn}{\nonumber}

\newcommand\myeq{\stackrel{\mathclap{\normalfont\tiny\mbox{$N \gg 1$}}}{\approx}}

\newcommand{\beq}{\begin{eqnarray}}
\newcommand{\eeq}{\end{eqnarray}}
\newcommand{\barray}{\begin{eqnarray}}
\newcommand{\earray}{\end{eqnarray}}
\newcommand{\disp}[1]{Eq.~(\ref{#1})}

\newcommand{\eps}{\epsilon}
 
\newcommand{\beg}{\begin{equation}}
\newcommand{\en}{\end{equation}}

 \newcommand{\lam}{\lambda}

\newcommand{\eref}[1]{Eq.~(\ref{#1})}
\newcommand{\re}[1]{(\ref{#1})}

\newcommand{\half}{\frac{1}{2}}
\newcommand{\E}{{\cal E}}
\renewcommand{\P}{{\bf \Gamma}}

\newcommand{\J}{{\cal J}}
\newcommand{\bgamma}{ {  \tilde{\bm \gamma}}}
\renewcommand{\S}{{\cal S}}

\begin{document}
\title{  Weyl's Relations,  Integrable Matrix Models and Quantum Computation }

\author{B. Sriram Shastry\footnote{sriram@physics.ucsc.edu}}
\affiliation{Physics Department, University of California, Santa Cruz, CA 95064, USA}
\author{Emil A. Yuzbashyan\footnote{eyuzbash@physics.rutgers.edu}}
\affiliation{Department of Physics and Astronomy, Center for Materials Theory, Rutgers University, Piscataway, NJ 08854, USA}
\author{Aniket Patra\footnote{patraaniket@gmail.com}}
\affiliation{Center for Theoretical Physics of Complex Systems, Institute for Basic Science, Daejeon 34126, Republic of Korea}

\date{\today}

\begin{abstract}
Starting from a generalization of Weyl's relations  in finite dimension $N$, we show that the Heisenberg commutation relations can be satisfied in a specific $N$-1 dimensional subspace, and display a linear map for projecting operators to this subspace. This setup is used to 
 construct a hierarchy of parameter-dependent commuting matrices in $N$ dimensions. This family of commuting matrices is then related to Type-1 matrices  representing quantum integrable models.  The commuting matrices find an interesting application in quantum computation, specifically in Grover’s database search problem. Each member of the hierarchy serves as a candidate Hamiltonian for quantum adiabatic evolution and, in some cases, achieves higher fidelity than standard choices --- thus offering improved performance.  
\end{abstract}


\maketitle

\date{\today}
\section{ Introduction}

In this work, we present a link between Weyl's relations, commuting matrices and   quantum computation, within the context of finite-dimensional matrices representing quantum systems in finite-dimensional Hilbert spaces.
To this end, we construct a class of mutually commuting real symmetric $N \times N$ matrices that depend linearly on a parameter $x$, representing quantum integrable systems in finite dimensions.
 These  are obtained  from an algebra involving  three matrices ($\E,\S,\P$), in a structure  that is closely related to generalizations  of  matrices appearing in Weyl's relations  \cite{Weyl,Schwinger,Schwinger2,Tekumalla}.  Weyl's concern was with the emergence of the Heisenberg commutation relations in the limit of large  dimensions.  We display variants of the Weyl matrices defining Weyls's relations in finite dimensions, which yield  canonical commutators in well defined subspaces defined by the orthogonality with respect to a single specific state. Using these, we construct a hierarchy of commuting matrices and find that they coincide with certain linear combinations of Type-1 commuting matrices previously obtained by us 
in~\cite{shastry,shastry1,haile,haile1,scaramazza,scaramazza1,yuzbashyan}, using entirely different arguments.

We show that the commuting matrices found here   generalize a model Hamiltonian studied in the context of
 quantum computation. It is used to display the quantum advantage over classical methods in the problem of a search in a random database --- i.e., Grover's algorithm as  
implemented through adiabatic evolution~\cite{Grover1, Grover2, Grover-Analysis, Farhi-Guttman, QAA_Theory, QAA, LAA, Berkeley, Childs, Young}. 

A detailed review of Type-1 matrices is available  
in~\cite{yuzbashyan}. These constitute a class of finite-dimensional matrices that aim to capture the essence of quantum integrability at the level of a matrix representation of models, rather than at the more conventional Hamiltonian level involving physical degrees of freedom.
  Here we derive the Type-1 commuting matrices  from the commutation relation~\re{comm0}.  

We first discuss the Weyl relations in the large $N$ limit in Section \ref{N-infinity}, where the Weyl matrices $A$ and $B$ [see \disp{A-matrix} and \disp{B-def}]   are introduced and  a review of how the Heisenberg algebra emerges from them is provided taking a careful continuum limit \cite{Schwinger,Tekumalla,comment-Santhanam}. In Section \ref{N-finite}, we summarize earlier generalizations of the Weyl's operator basis of \disp{U-Weyl}, which are used for creating a basis of operators representing maximally entangled states. We then introduce a matrix $C$ \disp{C-op1}, a new generalization of the Weyl matrices, which together with $B$ defines an algebra in \disp{Comm-1} that plays an important role in the subsequent analysis. This is a simplified version of the algebra found  in Section \ref{Intro-Weyl-3}  in \disp{comm0}, which is finally employed to generate the commuting matrices in Section \ref{Im}. In Section \ref{quantum-computation}, we discuss the use of the Type-1 matrices in implementing  the Grover algorithm  through adiabatic computation. By studying the fidelity, we show that the higher order matrices lead to an increase in the efficiency of computation. In Section \ref{Conclusions}, we make some final remarks.


\section{Weyl matrices and the limit $N\to \infty$}  \label{N-infinity}
Weyl proposed an important pathway for  understanding the Heisenberg algebra
\beq [Q,P]= i \hbar \iden \label{Heisenberg}, \eeq 
emerging  from a finite-dimensional setting, in the limit of infinite dimensions. Taking  the limit of infinite dimensions is necessary to avoid a familiar contradiction, arrived at by taking the trace of the Heisenberg algebra in finite dimensions. Weyl's work was followed by   others detailing and extending his  arguments \cite{Weyl,Schwinger,Schwinger2,Tekumalla}. Since our work relies on yet another extension of Weyl's work, before describing it we first provide a  quick summary of the known results.

 Weyl begins with a  relation  for a pair of  $N\times N$  unitary matrices $A$ and $B$, 
\beq
A B = B A e^{i \omega_0} \label{Weyl-1},
\eeq
where $\omega_0=  \frac{2 \pi}{N}$  and $e^{i \omega_0}$ is the $N^{th}$ root of unity with integer $N$ --- taken to infinity at the end. 
It follows from iteration that $A^N=\iden=B^N$. The form of the matrices is  determined as follows: first we set $B$ to be a diagonal with  diagonal entries $\{b_0,b_1,\ldots,b_{N-1}\}$, and $b_r=e^{i \omega_0 r} $ (thus  satisfying $B^N=\iden$), i.e.,  
\beq 
B= \sum_{n=0}^{N-1} b_n |n \rangle \langle n|. 
\label{B-def} 
\eeq
  On taking the  matrix elements of \disp{Weyl-1}, we get (for $0\leq \{i,j\}\leq N-1$) the conditions
\beq
A_{ i j} b_j= e^{i \omega_0} \;  b_i A_{i j}. \label{Weyl-2}
\eeq
Together with the  choice  $b_n=e^{ i \omega_0 n}$, this implies that the only non-zero elements of $A$ are of the form $A_{j,j+1}$, and  we may represent $A$ as
\beq
A= \sum_{n=0}^{N-1} |n-1\rangle \langle n|,  \label{A-matrix}
\eeq
 with periodic boundary conditions  $n\equiv n+N$. Thus  $A$ is the left-shift operator on a ring of sites of length $N$. $A$ is diagonalized  in the plane-wave basis 
 \beq
 |k_r\rangle= \frac{1}{\sqrt{N}} \sum_{n=0}^{N-1} e^ {i k_r n} |n\rangle \label{k-state},
 \eeq
 where $k_r = r  \, \omega_0$ and $0\leq r\leq N-1$,
 so that $A| k_r\rangle= e^{i k_r} |k_r\rangle$, and
 \beq
 A=  \sum_r e^{i k_r} |k_r\rangle \langle k_r |. \label{A-in-kspace}
 \eeq
 
 Consider first the limit of $N\to \infty$. Weyl shows that the Heisenberg algebra emerges from the Weyl relation~\re{Weyl-1} by expanding both sides for large $N$. This process  requires a definition of the position operator $\hat{Q}$ and the momentum operator $\hat{P}$, which are defined by 
 \beg
 A=e^{i \xi \hat{P}/\hbar},\quad B=e^{i \eta \hat{Q}},
 \en
 together with the requirement that $\xi,\eta$ are each of ${\cal O}(\frac{1}{\sqrt{N}})$.  With this, \disp{Weyl-1} vanishes to ${\cal O}(\frac{1}{N^2})$, provided
 \beq
 \xi \eta\, [\hat{Q},\hat{P}]= i \hbar \omega_0. \label{Weyl-2}
 \eeq
 This leads to the Heisenberg algebra \disp{Heisenberg}, provided 
 \beq
 \xi \eta= \omega_0.
 \eeq
We discuss the details of this derivation in Appendix \ref{Hesenberg_Derivation_Details}.

\section{Weyl matrices with N finite }\label{N-finite} 

Generalizations of Weyl's matrices $A,B$ for finite $N$ have recently received much interest in quantum computation.  These are useful as they form  a basis of $N^2$ operators representing maximally entangled  states \cite{genweyl1,genweyl2,genweyl3}, and are referred to as the Weyl operator basis. Each element of the $N^2$-dimensional operator basis is explicitly given by
\beq
U_{n,m}=\sum_{r=0}^{N-1} e^{i n k_r} |k_r\rangle \langle k_{r+m} |,\label{U-Weyl}
\eeq
where $n$ and $m$ range over $0,N-1$. These matrices  are extensions of the Weyl matrix $A$ given in the form~\re{A-in-kspace}.  

We turn to a  generalization of Weyl's operators in a different direction guided by the question: Is it possible to preserve \disp{Heisenberg} as an operator identity while keeping the dimension $N$ finite, when acting on a well-defined subspace?  One  is faced with a similar question when quantizing constrained field theories, where the constraint is satisfied only as an operator acting on the physical states. This leads us to define another class of matrices, where the analog of \disp{Heisenberg} is satisfied on a $(N-1)\times(N-1)$ subspace of the original $N\times N$ space of states.  

Towards this, we now define a matrix operator
$C$ through the formula
\beq
C= \half \sum_{n \neq m} \frac{|n\rangle \langle m|-|m \rangle \langle n|}{b_n-b_m}. \label{C-op1}
\eeq
 This matrix is closely related to $\log A$ of \disp{logA-20}, with the omission of the $b_m$ and $b_n$ in the numerator of the second term. 
Notice that $C$ in \eref{C-op1} and $B$ in \disp{B-def} are defined in terms of $b_n$ and the following property is independent of the choice of $b_n$.
The commutator of $C$ with  $B$ in \disp{B-def} is  computed as
\beq
[C,B] &=&  \sum_{n \neq m} (|n \rangle \langle m|) \\
&=&  \iden- N |\psi\rangle \langle \psi|,  \label{Comm-1}
\eeq
where the projection operator is constructed from the normalized ``flat state'' $|\psi\rangle$, which is defined as  
\beq
|\psi\rangle&=& \frac{1}{\sqrt{N}} \sum_n|n\rangle = |k_{r=0}\rangle, \label{Flatstate-1}
\eeq  
and is also the zero wave-vector ($k_r = 0$) state from \eref{k-state}. From \disp{Comm-1}, we see that 
\beq 
[C,B]\,|\Psi\rangle= |\Psi\rangle, \label{Canonical} 
\eeq 
for any state $|\Psi\rangle=\sum_{r \neq 0} \alpha_r |k_r\rangle$ for arbitrary $\alpha_r$. This implies that \eref{Canonical} remains valid as long as $|\Psi\rangle$ is  contained in the $(N-1)$-dimensional subspace spanned by the basis $|k_r\rangle$ with $r \neq 0$ --- i.e., the subspace orthogonal to the state $|k_{r=0}\rangle$.
 The trace of $[C,B]$ vanishes due to the contribution from the last term. In this sense, the conjugate $N\times N$  matrices  $C$ and $B$ come close to the canonical  infinite-dimensional Heisenberg algebra~\re{Heisenberg}.
 A particular advantage of the structure of $C$ is that the commutator~\re{Comm-1} is valid for an arbitrary choice of $b_n$ including real values. Below, we show the close connection of $B$ and $C$ with matrices that arise in the context of  the Type-1 integrable family, both using real values for the $b_n$.
 
 We note that the matrix $C$ can be written in the $k$ representation by using \disp{k-state} as
\beq
C= \frac{N}{2 \pi} \sum_{r=0}^{N-1} k_r |k_r\rangle \langle k_{r+1}| \label{C-in-kspace},
\eeq
which  is a  minimally off-diagonal version of the log-Weyl matrix $\log A$ from \disp{logA-2}. The algebra satisfied by $C$ seems to be quite rich.  We  quote one further result found by commuting the expressions in \disp{logA-1} with the one in \disp{C-in-kspace}:
\beq
[\log A, C]= - i C. \label{Ehrenfest}
\eeq
This result  is reminiscent of the Fock  commutator of $\left[p, \half (px + x p)\right]=-  i \hbar$, which is used to obtain the virial theorem in quantum theory. The commutator $[\log B, C]$ is not related simply to $C$, so the analogy is only partial.  We also note  that the matrix $C$ in \disp{C-in-kspace} is  similar to matrix  $U_{1 1}$  of \disp{U-Weyl}, and is obtained, up to an overall prefactor, when  the $e^{i k_r }$  is replaced by $i k_r $.

In summary, the three  matrices: $C$ in \disp{C-op1} and $A$ and $B$ in \disp{Weyl-1}, and their logarithms introduced in \disp{logA-20} and \disp{logB},  satisfy the basic commutation results~\re{Comm-1} and \re{Ehrenfest} for the special choice of $b_n=e^{i \omega_0 n}$. These relations, with the exception of \disp{Ehrenfest}, generalize to arbitrary choices of $b_n$. We may qualitatively view $B$ as the position variable itself, and then $A$ can be  viewed as the discrete translation operator. The role of $C$ is somewhat context driven. In  \disp{Ehrenfest} $C$ is analogous to the Fock operator $xp+px$, but in  \disp{Canonical}, C is analogous to the momentum operator.   In  \disp{Canonical}  we saw  that $C$ and $B$ are  canonically conjugate  in a subspace which excludes the state $|\psi\rangle$ of \disp{Flatstate-1}. Thus, transforming arbitrary operators so that they annihilate the flat state is expected to be very useful --- and we follow up on this idea below. We will refer to $C$ and $B$ as {\em generalized Weyl matrices} when the entries $b_n$ are chosen arbitrarily.

We also mention the work on   a truncated harmonic oscillator restricted to operate in an $N$-dimensional  Hilbert space \cite{Buchdahl,Buchdahl-further1, Buchdahl-further2}, which leads to a similar algebra as \disp{Comm-1}, but  with fixed expressions for the matrices.

\section{The algebra of Type-1 matrices and connection with generalized Weyl matrices  \label{Intro-Weyl-3}}

A brief summary of the Type-1 matrix models is provided next, interested readers can find more details in the review \cite{yuzbashyan}. In \cite{shastry,shastry1,haile,haile1,yuzbashyan,scaramazza}, we developed a theory of sets of real symmetric $N\times N$ matrices depending linearly on a real parameter, say $x$, as finite dimensional prototypes of quantum integrable systems. These emerge from a study of the conservation laws of quantum integrable systems, such as the 1-d Heisenberg and 1-d Hubbard models, when restricted to finite sizes. The matrix formulation has the advantage of being ``blind'' to the originating quantum  model, while capturing elements related to their quantum integrability. 
Members of the commuting set are  in the form $\alpha(x)=a + x A$, where $a$ and $A$ are $N\times N$ real symmetric matrices obeying certain constraint relations.  A convenient  formulation to define this set of matrices is to introduce $N$ objects $Z_j$ that play the role of a basis:
\beq
Z_j = |j \rangle \langle j| + x \sum_{k\neq j} \frac{1}{\eps_j-\eps_k} \left\{\; \gamma_j \gamma_k ( |j \rangle \langle k| +|k \rangle \langle j|) - \gamma_j^2 |k \rangle \langle k| -\gamma_k^2 |j \rangle \langle j|   
\; \right\} \label{Z-s},
\eeq
where the commutation  $[Z_j,Z_l]=0$ for all $j,l$ is established, with an  unconstrained  set of $2 N+1$ parameters $\{\gamma_j,\eps_j\}$ and $x$. We may then write $\alpha(x)=\sum_j c_j Z_j$, and varying the coefficients $c_j$ generates the family of mutually commuting Type-1 matrices.

In this  program of constructing finite-dimensional matrices representing  quantum integrable systems, we  unexpectedly encountered an algebraic structure~\cite{scaramazza} that bears considerable resemblance to the above problem~\re{Comm-1}. We show here that our integrable matrix model, and  its conservation laws can be obtained purely from this algebra. This calculation thereby offers a new perspective on our model and provides further insights.

Our main cast consists of a set of  matrix operators $\S$, $\E$, $\P$, and ${\bf D}$  with real entries acting on the $N$-dimensional real vector space ${\cal R}_N$. The antisymmetric matrix $\S$ is defined through its matrix elements
 \beq
 \S_{ij} = x \; (1-\delta_{ij})  \frac{\gamma_i \gamma_j}{\eps_i-\eps_j}, \;\; \mbox{ as   } \S= \sum_{i,j} | i\rangle\, \S_{ij}\, \langle j|  \; ,\label{S-form}
 \eeq
 where $x$ is a real parameter, $\{\eps_1,\eps_2,\ldots, \eps_N\}$ are $N$ real variables, and $\{\gamma_1,\gamma_2,\ldots,\gamma_N\}$ are $N$ real numbers normalized by the condition 
 \beg
 \sum_{i=1}^N\gamma_i^2=1.
 \en
 We also need three diagonal matrices:
 \beq
 \E= \sum_{j=1}^N \eps_j |j \rangle \langle  j|, \label{E-form}
 \eeq
 \beq
{{\bf D}} = \sum_{j=1}^N \gamma_j^2 | j \rangle \langle j|. \label{D-form}
\eeq 
and a projection operator
\beq
\P& = & |\gamma \rangle \langle \gamma|, \;\;\mbox{where} \label{Gamma-def}\\
 | \gamma \rangle & =& \sum_{j=1}^N \gamma_j | j \rangle. \label{gamma-def} \eeq
It follows from  the normalization of $\{\gamma_i\}$ that  $\mbox{Tr} \, {\P}=1$ and  $\mbox{Tr} \, {\bf D}=1$.

It is seen by comparing \disp{S-form} with \disp{C-op1} that ${\cal S}$ is  the same as $C$  introduced earlier, when the constants $\gamma_i\to 1$, and $ \eps_i \to b_i $. With these changes, we see that ${\cal E}$ becomes $B$ of \disp{B-def}  and $\bf D$ reduces to $\iden$.  Finally, we see that $\P$ is proportional to $|\psi\rangle \langle \psi|$ where $|\psi\rangle$ is given in \disp{Flatstate-1}, and the flat states $|\gamma\rangle$ map onto $|\psi\rangle$. With these changes, the algebra discussed below can be viewed as a generalization of Section \ref{N-finite}.

The   commutator of $\E$ and $\S$ is easily seen to be  
\beq
[\E,\S]=  \, x \,  \left(    \P -  {{\bf D}}   \right), \label{comm0}
\eeq
with a vanishing trace of both sides, as expected.
In the special case of $\gamma_i=\frac{1}{\sqrt{N}}$ for all $i$, we see that \disp{comm0} reduces to \disp{Comm-1}, and  $x$ is analogous to the Planck's constant.

Below   we uncover  $N$ mutually commuting operators $I_n $ [see \disp{our-Im}] using the three building blocks $\E, \P, \S$. Among these, there is another natural candidate,  \disp{our-H}, that can be viewed as the Hamiltonian. Indeed  we will  see later that \disp{our-H}  is a natural Hamiltonian for a Grover type quantum search algorithm, as discussed in \cite{Farhi-Guttman,QAA,QAA_Theory,LAA}.

\subsection{  A linear map on operators \label{linear-map}}
Guided by our earlier discussion [see paragraphs below \disp{Ehrenfest}], we now find a systematic method for mapping any operator to a related one which annihilates the projection operator  $\P$. These  are found by  first analyzing the action of arbitrary operators $\Q$ on the state $|\gamma\rangle$.
 Guided by the form of \disp{gamma-def}, we now introduce  a useful decomposition for any symmetric operator $\Q$ 
 \beq 
 \Q=\Q_{||}+\Q_\perp, 
 \eeq
 where  $\Q= \sum_{ij}| i\rangle Q_{ij} \langle j|$ and $Q_{ij}=Q_{ji}$. The two pieces of $\Q$ have the property that
  \beq \Q |\gamma \rangle =\Q_{||} |\gamma\rangle\;\mbox{  and  } \Q_{\perp}|\gamma \rangle=0. \label{perp2}\eeq
More explicitly,
\beq
\Q_{||}= \sum_{ij} |i \rangle \langle i| \;  Q_{ij} \left( \frac{\gamma_j}{\gamma_i} \right),
\label{parallel}
\eeq 
and
\beq
\Q_{\perp}= - \half  \sum_{ij} \frac{1}{\gamma_i \gamma_j } \Bigl\{ |i \rangle \gamma_j - |j \rangle \gamma_i \Bigr\} \;Q_{ij} \; \Bigl\{  \gamma_j \langle i|  - \gamma_i \langle j|  \Bigr\}. \label{perpendicular}
\eeq

 It is clear that $\Q_{\perp}$  annihilates  the vector $|\gamma\rangle$,
because  the terms  $\Bigl\{  \gamma_j \langle i|  - \gamma_i \langle j|  \Bigr\}$ annihilate the state $|\gamma\rangle$ for every pair $i,j$.
Using the symmetry of $\Q$, \disp{perpendicular}  also implies the orthogonality condition
\beq
\Q_{\perp}. \P =0= \P. \Q_{\perp}. \label{orthogonality}
\eeq

 This construction can also be viewed as a  linear map $ \Q\to \Q_{||} $. Towards this end, we define the equal amplitude (flat sum) state $|\Phi\rangle$:  
\beq
|\Phi \rangle = \sum_j | j \rangle.  \label{flat}
\eeq
It is useful to define an invertible diagonal operator
\beq
\bgamma&=& \sum_j \gamma_j | j \rangle \langle j |, \;\;\;\;\mbox{ so that } \nn \\
| \gamma \rangle  &=& \bgamma |\Phi\rangle. \label{gamma-state}
\eeq

In terms of $ \bgamma $ it is easy to see that 
\beq
&&  \Q_{||} \equiv \sum_j \;\; \langle j | \bgamma^{-1} \Q \bgamma | \Phi \rangle \; |j\rangle \langle j | \label{linmap} \label{parallel2}.
\eeq

In summary, the linear operator  map~\re{linmap}  helps us to decompose  an arbitrary  real symmetric operator $\Q$ into two  components, $\Q=\Q_{||}+ \Q_{\perp}$, where the latter annihilates the state $|\gamma\rangle$.

\subsection{Analogy with stochastic equations \label{stochast} }

A simple analogy  between the construction in Eqs.~(\ref{parallel}), \re{perpendicular}, and \re{parallel2},  and  stochastic  time evolution equations in nonequilibrium physics  may be  helpful here.  In the study of master equations,  the equilibrium state can be used to perform a similarity transformation that enables the   interpretation of the  transition operator as a Hamiltonian,  and the master equation itself as    the Euclidean version of  Schr\"odinger's equation \cite{MasterEquation1, MasterEquation2, MasterEquation3}.  If we consider  a  state vector $|P\rangle$, with  components representing the probabilities of the various basis states,  and a transition operator ${\cal T}$  satisfying  a master equation
\beq
\partial_t |P \rangle = - {\cal T} |P \rangle \label{master},\;\;\;\; \langle \Phi | {\cal T}=0, 
\eeq
where $|\Phi\rangle$  defined in \disp{flat} gives equal weight to every configuration $|i\rangle$,  then the total  probability is conserved. We can write the equilibrium state as
\beq
|{P_{eq}} \rangle = {\bgamma}^2 |\Phi\rangle = \sum_j \gamma_j^2 |j \rangle, \label{equibilibrium}
\eeq 
with positive weights for each configuration, so that the system flows to equilibrium provided
$
{\cal T} |P_{eq} \rangle=0.
$

In order to transform this to the Hamiltonian formulation one  performs a similarity transformation
\beq
|P\rangle = \bgamma |\Psi\rangle, \;\; H= \bgamma^{-1} {\cal T} \bgamma, \;\;
\eeq 
giving rise to the imaginary time Schr\"oedinger equation
\beq
\partial_t | \Psi \rangle= - H | \Psi \rangle. \;\;
\eeq
The equilibrium  distribution under this similarity transformation  gives rise to the ground state wave function
 $|\Psi_{0}\rangle\equiv \bgamma ^{-1} |P_{eq}\rangle = \bgamma |\Phi\rangle$, which satisfies the conditions
 \beq
 H |\Psi_0\rangle = 0 = \langle \Psi_0| H.
 \eeq
For a class of  transition matrices this prescription gives a Hermitian Hamiltonian.  The state $|\gamma\rangle$ in \disp{gamma-state} now plays the  role of the ground state $|\Psi_{0}\rangle$.  This is the route taken in connecting  Dyson's Brownian motion of matrices with the Calogero-Sutherland model  of interacting particles in one dimension.

 \section{ The parameter-dependent matrix Hamiltonian and its conservation laws \label{Im}}
 
Using the above algebra, we  construct a Hamiltonian $H$  as
\beq
H=  \E + x (\P-\iden) \; \equiv I_1 \label{our-H}
\eeq
We  also refer to  $H$ as $I_1$, since, as we show below, it belongs to a family of matrices,
\beq
I_m= \E^m +  [\E^m,\S]_{\perp},
\label{perp} 
\eeq
that commute among each other.

From \disp{comm0} we note that the commutator  $[\E^m,\S]$  is linear in $x$. 
We thus define
\beq
  [\E^m,\S]_\bot  \equiv     x\, K_m,  \label{def-Km}
\eeq
so that the 
 the conservation laws~\re{perp}  now read
 \beq
I_m = \E^m + x \ K_m,    \label{our-Im}
\eeq
 with  $m=1,2,\ldots N$. 
Here the effective Planck's constant $x$ is arbitrary, but common to all the members of this commuting family. Similar constants arise in most quantum integrable systems, e.g., the $U$ parameter in the 1-d Hubbard model.
These satisfy for all $n,m \leq N$
\beq
[H,I_m]=0,  \label{hcom} \\
~[I_n,I_m]=0. \label{Icom}
\eeq
In fact, we will begin by showing that the first member of the sequence in \disp{our-Im} is  the Hamiltonian~\re{our-H}.  For this purpose we use \disp{comm0} to write
\beq
[\E,\S]_{\perp}= x (\P- {{\bf D}})_{\perp} = x ( \P- \iden),
\eeq
where we used ${{\bf D}}_{\perp}=0$, a result   valid for any diagonal operator, thereby proving that $I_1=H$. It may be useful to   record the explicit form of the next conservation law 
\beq
I_2= \E^2+ x \left( \E \P+ \P \E -\E - \iden \; \langle \gamma | \E | \gamma \rangle \right).
\eeq
 Let us now prove \disp{hcom} for general $m$. The commutator is quadratic in $x$,  hence we need the vanishing of the three terms $x^\alpha$, with $\alpha=0,1,2$ separately. The constant term is trivial since $[\E,\E^m]=0$. The $O(x)$ term requires
\beq
 [\E^m,[\E,\S]_\bot]=  [\E, [\E^m,\S]_\bot ]. \label{req1} \eeq
We note that the perpendicularity condition can be dropped on both sides, since the parallel part commutes with $\E$ and $\E^m$.  We invoke the Jacobi identity 
\beq
 [\E^m,[\E,\S]]+[\S,[\E^m,\E]]+ [\E,[\S,\E^m]]=0,
\eeq
and since the middle term is identically zero, this implies \disp{req1}. The $O(x^2)$ term requires
\beq
\P . [\E^m,\S]_{\perp} = [\E^m,\S]_{\perp} . \P. 
\eeq
Now both sides vanish by the definition~\re{orthogonality} of the orthogonality, and thereby we have proved \disp{hcom} for any integer $m$.

\subsection{ Mutual  Commutation of $I_n$} 

 If we assume  that the spectrum of $H$ is nondegenerate, the commutation of the operators~\re{Icom} is implied by a standard argument from linear algebra.  However it is worthwhile proving \disp{Icom} directly without taking recourse to making assumptions about the spectrum of $H$. Note that we have proved \disp{hcom} purely algebraically, without enquiring into the form of $\S$ and ${{\bf D}}$.  It might be possible to do so for \disp{Icom} as well, but we now present a proof that uses the explicit solution  for $\S$ and ${{\bf D}}$.

Returning to \disp{perp} and using the explicit expressions~(\ref{S-form}), (\ref{E-form}), and (\ref{D-form}), we first work out
\beq
\frac{1}{x} \; [\E^m,S]_{ij}= \gamma_i \gamma_j \frac{\eps_i^m-\eps_j^m}{\eps_i-\eps_j}.
\eeq
It is useful to introduce the vectors with circular brackets
\beq
|p) = \sum_j \gamma_j \eps_j^p |j \rangle, \label{roundstates}
\eeq
so that $|0)= |\gamma \rangle$. We further define the overlap  and projector $\J$
\beq
(p|q)&=& \sigma_{p+q}, \;\; \J(p,q)= |p)(q|, \nn \\
 \mbox{where}\; \sigma_p &=& \sum \gamma_i^2 \eps_i^p.
\eeq
Thanks to the normalization, we have $\sigma_0=1$. In terms of these we find
\beq
\frac{1}{x} \;  [\E^m,S]= \sum_{r=0}^{m-1} \J(m-1-r, r) - m \sum_j \gamma_j^2 \eps_j^{m-1} |j\rangle \langle j|.
\eeq
We can construct the perpendicular parts of these operators immediately from the definitions:
In terms of a set of diagonal operators
\beq
\E^p = \sum_j \eps_j^p |j\rangle \langle j| \label{Ds}
\eeq
we obtain
\beq
\frac{1}{x} \;  [\E^m,S]_\bot = K_m=  \sum_{r=0}^{m-1} \left( \J(m-1-r, r) - \E^{m-1-r} \; \sigma_r \right), \label{Emperp}
\eeq
so that 
\beq
I_m=\E^m+ x \ \sum_{r=0}^{m-1} \left( \J(m-1-r, r) - \E^{m-1-r} \; \sigma_r \right).
\eeq

With this explicit form for $I_m$, we compute the commutator
\beq
[I_m,I_n]=0,
\eeq
the $O(x)$ term vanishes since  we can ignore the $\bot$ constraint under the commutator with diagonal operators:
\beq
[\E^m,[\E^n,S]_\bot ] - [\E^n,[\E^m,S]_\bot ] = [\E^m,[\E^n,S] ] - [\E^n,[\E^m,S] ] .
\eeq
The right-hand side    vanishes upon applying the Jacobi identity. Thus the only non trivial condition to check is the $O(x^2)$ term  $[K_m,K_n]=0$, so we write
\barray
[K_m,K_n] & =& \sum_{r=0}^{m-1} \sum_{s=0}^{n-1} [\left( \J(m-1-r, r) - \E^{m-1-r} \; \sigma_r \right), \left( \J(n-1-s, s) - \E^{n-1-s} \; \sigma_s \right)],\nn \\
&=& \sum_{r=0}^{m-1} \sum_{s=0}^{n-1} \left( \sigma_{\mu} \ \bar{\J}(r,s)  + \sigma_r \ \bar{\J}(s,\mu) + \sigma_s \ \bar{\J}(\mu,r) \right), \label{kscommute}
\earray
where we  abbreviated $\bar{\J}(a,b) = \J(a,b)-\J(b,a)$,  $\mu=m+n-2-r-s$, and  used $\E^p |q)= |p+q)$. The term vanishes upon performing the sum, as we next show.

\subsection{ Proof of cancellation of \disp{kscommute}:}

Let us write $m\to m+1$ and $n\to n+1$ for convenience and write the $(-1\times)$ right-hand side of \disp{kscommute} in the form
\beq
{\cal Y}= \sum_{r=0}^m \sum_{s=0}^n \{ s,r | m+n-r-s \},
\eeq
where we denote arbitrary integers by $a,b,c$ and the symbol $\{ \}$ stands for the cyclic sum
\beq
\{a,b |c \}&=& [a,b;c]+[c,a;b]+[b,c;a], \nn \\
~[a,b;c]&=& \bar{\J}(a,b) \sigma_c.
\eeq
In fact, the detailed form of $[a,b;c]$ is not important. The only property needed is the skew symmetry
\beq
~[a,b;c]=-[b,a;c].
\eeq
From these  definitions it is easily  seen that 
\beq
\{a,b|c\}&=&\{c,a|b\} = \{b,c|a\},  \;\;\;\;\mathrm{(i)} \nn \\
\{a,b|c\}&=& - \{b,a|c\}, \;\;\;\;\;\;\;\;\;\;\;\;\;\;\;\;\mathrm{(ii)}\nn \\
\{a,a|c\}&=&0. \;\;\;\;\;\;\;\;\;\;\;\;\;\;\;\;\;\;\;\;\;\;\;\;\;\;\;\;\mathrm{(iii)} \label{props1}
\eeq 
We now separate ${\cal Y}$ into two parts assuming $m > n$,
\beq
{\cal Y}&=& {\cal Y}_1+ {\cal Y}_2, \nn \\
{\cal Y}_1&=&\sum_{r=0}^n \sum_{s=0}^n \{ s,r | m+n-r-s \}, \nn \\
{\cal Y}_2&=&\sum_{r=n+1}^m \sum_{s=0}^n \{ s,r | m+n-r-s \}.
\eeq
We first note that ${\cal Y}_1$ vanishes since the ranges of $s$ and $r$ are identical, so we can switch them, $r \leftrightarrow s$, and use property (ii) in \disp{props1},  so that ${\cal Y}_1=-{\cal Y}_1=0$. To analyze ${\cal Y}_2$, we fix $r$ in its new range, and observe that the resulting   range of the integer $(m+n-r-s)$ coincides with that of $s$. For a given $r$ we can now see that as $s$ varies, there are two cases:  (a) the  integer $(m+n-r-s)$ equals $s$, which vanishes by using property (iii) in \disp{props1}, or case (b) integer $(m+n-r-s)$ is distinct from $s$, in which case we can exchange these,   $(m+n-r-s) \leftrightarrow s $, and then use property (ii) in \disp{props1} to show that the sum of these terms vanishes. 

We have thus shown that the algebra~\re{comm0} directly leads to the  conservation laws $I_n$. We can take the linear sums $\sum_j c_j I_j$ and these are still constants of motion.  The earlier construction of the Type-1 family can be related to the $I_n$ straightforwardly. Indeed,  we can  express the operators $I_m$ of \disp{our-Im} in terms of  $Z_j$ given by \eref{Z-s} as
\beq
I_n = \sum_j \eps_j^n Z_j. 
\label{Im-as-Z}
\eeq
Note that this relationship implies that only $N-1$ of the infinite sequence of $I_n$ are linearly independent.

\subsection{ Spectra, recurrence relation, general formula, and $I_n$ as a polynomial in  $H$:}

  The spectrum of a general type-1 matrix   was determined in~\cite{haile}. The eigenvalues $\xi_\alpha$ and unnormalized eigenvectors
  $|\varphi_\alpha\rangle$ (shared by all matrices in the commuting family) are
  \beg
  \xi_\alpha= x\sum_{i=1}^N \frac{d_i\gamma_i^2}{\lam_\alpha - \eps_i},\quad \varphi^{(i)}_\alpha=\frac{\gamma_i}{\lam_\alpha-\eps_i},\quad \alpha=1,\dots,N.
  \label{sp}
  \en
  where $d_i$ are the eigenvalues of the $x=0$ part of the matrix and $\lam_\alpha$  are the $N$ roots of the equation
  \beg
  \sum_{i=1}^N \frac{\gamma_i^2}{\lam_\alpha-\eps_i}=\frac{1}{x}.
  \label{lambda}
  \en
  For $I_n$ we have $d_i=\eps_i^n$,  i.e., the eigenvalues of $I_n$ read
 \beg
 \eta^{(n)}_\alpha=x\sum_{i=1}^N \frac{\eps_i^n\gamma_i^2}{\lam_\alpha - \eps_i}.
 \label{eta}
 \en
 Note that when $x\to 0$, $\lam_\alpha\to \eps_i$ and $\eta^{(n)}_\alpha\to \eps_i^n$.   When $x\to\infty$, one $\lambda_\alpha$ (say $\lambda_N$) diverges, while others remain finite. \eref{lambda} implies  \beg
 \lambda_N\to x\sum_i\gamma_i^2=x.
 \en
 The corresponding eigenvector (upon normalization) $|\varphi_N\rangle\to|\gamma\rangle$ and the eigenvalue $\eta^{(n)}_N\to 0$.  Indeed, Eqs.~\re{perp2} and \re{def-Km} imply that $|\gamma\rangle$ is an eigenvector of $I_n$ in $x\to\infty$ limit with eigenvalue 0.  
 
 Eigenvalues of $I_n$ are polynomials in $\lambda_\alpha$ of order $n$. Indeed, substituting the identity 
 \beg
 \eps_i^n=(\eps_i- \lam_\alpha)(\eps_i^{n-1}+\eps_i^{n-2}\lam_\alpha +\dots +\eps_i\lam^{n-1}_\alpha+\lam^{n-1}_\alpha)+\lam^n_\alpha
 \en
 into \eref{eta} and using \eref{lambda}, we derive
 \beg
 \eta^{(n)}_\alpha=\lam^n_\alpha-x\sum_{k=1}^n a_k\lam^{n-k}_\alpha,\quad a_k=\sum_{i=1}^N \eps_i^{k-1}\gamma_i^2=\langle\gamma|\E^{k-1}|\gamma\rangle,
 \label{poly1}
 \en
 which also implies a recurrence relation for eigenvalues of $I_n$
 \beg
 \eta^{(n+1)}_\alpha=\lam_\alpha \eta^{(n)}_\alpha-x a_n.
 \label{rec}
 \en
 Note in particular that the eigenvalues of $H$ are 
 \beg
 \eta^{(1)}_\alpha=\lam_\alpha-x\sum_i\gamma_i^2=\lam_\alpha-x.
 \en
 
 The above expressions for the eigenvalues are helpful in establishing several facts. For example, $I_n$ is a polynomial in $H$ of order $n$. To write down this polynomial, observe that
 the eigenvalues of $H+x\iden=\E+x\P$ are simply $\lam_\alpha$, and Eqs.~\re{poly1} and \re{rec} therefore mean
 \beg
 I_n=(H+x\iden)^n -x\sum_{k=1}^n \langle\gamma|\E^{k-1}|\gamma\rangle (H+x\iden)^{n-k}, \quad I_{n+1}=HI_n+x I_n - x\langle\gamma|\E^{n}|\gamma\rangle.
 \label{in}
 \en
 Further, replacing $H+x\iden$ with $\E+x\P$ in this equation and tracking down the terms linear in $x$ (higher orders in $x$ cancel identically), we derive $I_n$ explicitly in terms of $\E$ and $\P$,
 \beg
 I_n=\E^n+xK_n=\E^n+x\sum_{k=1}^n\left(\E^{n-k}\P\E^{k-1}-\E^{n-k}\langle\gamma|\E^{k-1}|\gamma\rangle\right).
 \label{CC_Type-1_Relation}
 \en

 Finally, let us obtain the eigenvalues $\kappa^{(n)}_\alpha$ of $K_n$. Evidently,
 \beg
 \kappa^{(n)}_\alpha= x^{-1}\lim_{x\to\infty} \eta^{(n)}_\alpha.
 \en
 As discussed above, in this limit $\lam_N\to x\to\infty$. Remaining $\lam_\alpha\to \tilde\lam_\alpha$, where $ \tilde\lam_\alpha$ are finite and are the   roots of the equation 
 \beg
  \sum_{i=1}^N \frac{\gamma_i^2}{\tilde\lam_\alpha-\eps_i}=0,\quad \alpha=1,\dots,N-1.
  \label{lambdat}
  \en
Plotting the left-hand-side of this equation as a function of $\tilde\lam_\alpha$, we observe that the roots are sandwiched between $\eps_i$. Assuming $\eps_i$ are in ascending order, we have
\beg
\eps_1< \tilde\lam_1 < \eps_2< \tilde\lam_2 <\dots < \eps_{N-1} < \tilde \lam_{N-1}<\eps_N.
\label{order}
\en
Further, \eref{eta} implies  $\kappa^{(n)}_N=0$ (these zero eigenvalues correspond to the common eigenvector $|\gamma\rangle$ of $K_n$), while \eref{poly} implies
  \beg
 \kappa^{(n)}_\alpha= -\sum_{k=1}^n \langle\gamma|\E^{k-1}|\gamma\rangle \tilde \lam^{m-k}_\alpha,\quad \alpha=1,\dots,N-1.
 \label{poly}
 \en
Suppose all eigenvalues $\eps_i$ of $\E$ are nonnegative and at least one of them is nonzero, i.e., $\E$ is a positive semi-definite matrix of nonzero rank. Then, Eqs.~\re{order} and \re{poly} imply that $\kappa^{(n)}_\alpha<0$ for $\alpha=1,\dots,N-1$.  It follows that $|\gamma\rangle$ is the ground state of $-K_n$ with eigenvalue~0.

\section{Quantum  Computation with Type-1 Matrices} \label{quantum-computation}

We now explore the potential application of Type-1 integrable matrices in quantum search algorithms. In the classical setting, locating a marked item in an unstructured database of size~$N$ requires, on average, $N/2$ queries. Grover's quantum algorithm~\cite{Grover1, Grover2} offers a quadratic speedup, reducing the query complexity to $\mathcal{O}(\sqrt{N})$. We consider the adiabatic quantum computing implementation of this algorithm~\cite{Farhi-Guttman, QAA_Theory, QAA, LAA}, in which the system evolves from a uniform superposition of all basis states to the target state under the specific time-dependent Hamiltonian $H_G(t)$ (referred to here as the Grover Hamiltonian). Remarkably, it turns out  that $H_G(t)$   is Type-1 matrix---namely, the Hamiltonian defined in \eref{our-H}.   We then demonstrate how the commuting operators $I_n$ can be exploited to enhance the fidelity of the quantum computation.

In quantum search algorithms~\cite{Grover1, Grover2, Grover-Analysis, QAA, QAA_Theory, Young, Farhi-Guttman, Childs, Berkeley, LAA}, database items are represented as orthonormal basis states in a Hilbert space. The computation begins in the uniform superposition~\cite{gamma}
\beg
|\gamma\rangle = \frac{1}{\sqrt{N}}\sum_{x=1}^{N} |x\rangle,
\label{UniformSup}
\en
which equally weights all possible configurations. The goal is to identify a specific, unknown target state $|m\rangle$. Grover's algorithm achieves this by iteratively applying a sequence of unitary operations approximately $k \sim \pi \sqrt{N}/4$ times~\cite{Grover1, Grover2}.

In contrast to this discrete-time protocol, we focus on an analog version of the quantum search~\cite{Farhi-Guttman, QAA_Theory, QAA, LAA}. The system evolves under the time-dependent Grover Hamiltonian
\beg
\begin{split}
H_G(t) = s(t) \Bigl[ \iden - |m \rangle \langle m| \Bigr] + \left[ 1 - s(t) \right] \Bigl[ \iden - |\gamma \rangle \langle \gamma| \Bigr],
\label{Ham_RC_LocalAdiabatic}
\end{split}
\en
with interpolation function $s(t)$ satisfying $s(0) = 0$ and $s(T_\mathrm{run}) = 1$. In a perfectly adiabatic evolution, the system would follow the instantaneous ground state from $|\gamma\rangle$ at $t = 0$ to the target state $|m\rangle$ at $t = T_\mathrm{run}$. Achieving this ideal evolution would require an infinitely long runtime, whereas the practical objective is to minimize $T_\mathrm{run}$ through a suitable choice of $s(t)$, while ensuring the final state remains sufficiently close to $|m\rangle$.

Roland and Cerf~\cite{LAA} showed that requiring the fidelity $F(t)$ between the evolving state $\Psi(t)$ and the instantaneous ground state $\Psi_0(t)$ to remain above a fixed threshold,
\beg
F(t) = \left| \langle \Psi_0(t) | \Psi(t) \rangle \right|^2 \ge F_{\min},
\en
throughout the evolution under the Hamiltonian~\re{Ham_RC_LocalAdiabatic}---that is, ensuring local adiabaticity---recovers the quadratic speedup of Grover's algorithm. Assuming $F_{\min}$ is close to 1 and applying the adiabatic theorem, they further derive the corresponding interpolation function $s(t)$ as
\beg
\begin{gathered}
s(t) =  \frac{ N\tan\left(
\frac{2 t \delta}{N} \sqrt{N-1} \right)}{2\sqrt{N-1}\left[1 + \sqrt{N-1} \tan\left( \frac{2 t \delta}{N} \sqrt{N-1} \right)\right]},
\label{Parametric_Interpol_LocalAdiabatic}
\end{gathered}
\en
where $\delta = \sqrt{1 - F_{\min}}$.   The complexity of this locally adiabatic evolution governed by $H_G(t)$ is then determined by the condition $s(T_\mathrm{run}) = 1$, which yields
\beg 
  T_\mathrm{run} = \frac{1}{\delta}\frac{N}{\sqrt{N-1}}\arctan{\sqrt{N-1}} \;\;\; \myeq \;\;\; \frac{\pi}{2\delta}\sqrt{N}.
\label{Quad_LocalAdiabatic}
\en 
  
  We observe that the Grover Hamiltonian $H_G(t)$ is a special case of the Type-1 Hamiltonian $H$. To see this, note that the Hamiltonian $H$ defined in \eref{our-H}, along with its commuting partners $I_n$ in \eref{our-Im}, can be viewed as a two-parameter family of operators,
\beg
H = u\, \E + v\,  (\iden-\P), \quad I_n = u \, \E^n - v\,  K_n,
\label{2par}
\en
where we have redefined $x = -v/u$ and rescaled all operators by an overall factor of $u$. This reparameterization leaves the commutation relations $[H, I_n] = [I_k, I_n] = 0$ unchanged. We now fix the basis in which $\E$ is diagonal so that $|1\rangle = |m\rangle$, and choose the parameters
\beg
\eps_1 = 0, \quad
\eps_2 = \cdots = \eps_N = 1.
\label{Quant_Search_Param_Choice}
\en
With this choice, $\E = \iden - |m \rangle \langle m|$, and since $\P = |\gamma \rangle \langle \gamma|$ by definition, we find that $H = H_G(t)$ along the one-parameter trajectory $u = s(t)$, $v = 1-s(t) $ in the $(u, v)$ parameter space. The fact that $H_G(t)$ lies within the Type-1 family is notable, given that Type-1 matrices form a measure-zero subset of all matrices of the form $M(u,v) = u A + v B$, where $A$ and $B$ are arbitrary real symmetric matrices independent of    $u$ and $v$~\cite{haile,haile1}.

\begin{figure}
    \centering
    \includegraphics[trim={0.75cm 0.0cm 0.7cm 0.0cm},clip,scale=0.75]{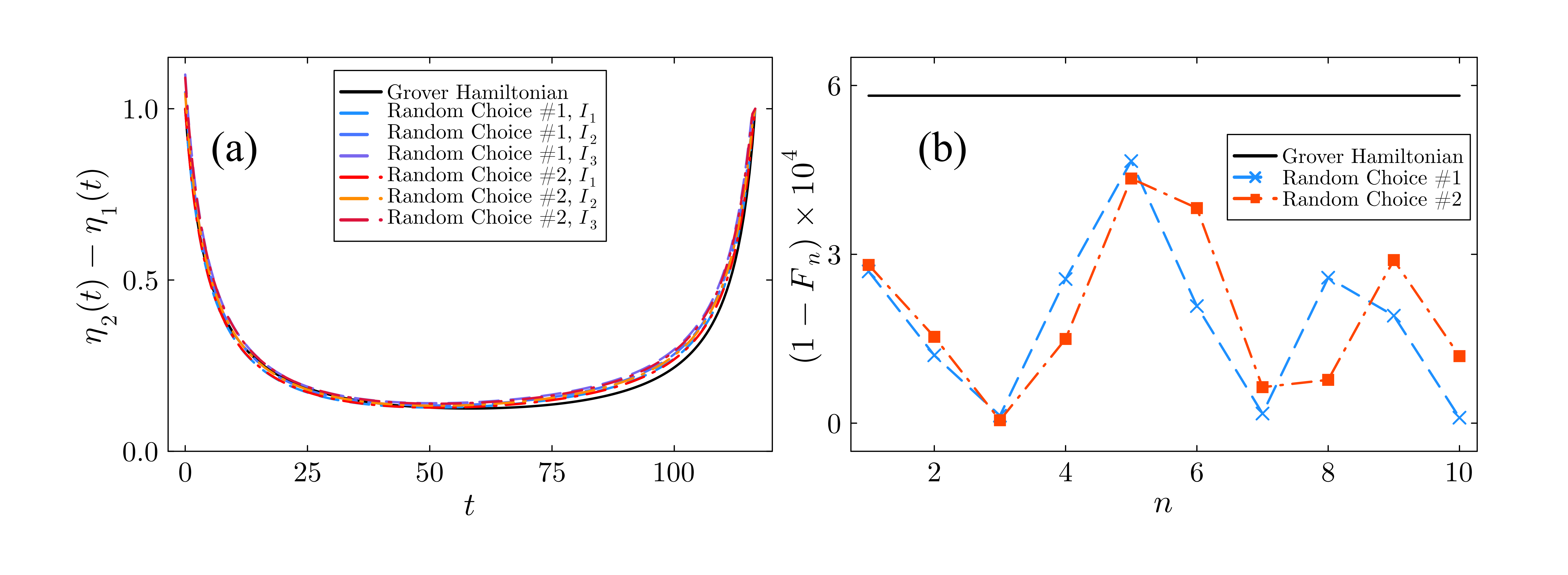}
    \caption{ Efficiency increase of the Grover's search algorithm by utilizing quantum integrability of Type-1 matrices. Panel (a): Instantaneous energy gaps between the ground and first excited state as  functions of $t\in (0, T_\mathrm{run})$ for   two random sets of $\eps_{i>2}$ with $\Delta\eps = 0.1$ and three different currents $I_{n}$ as well as for the Grover Hamiltonian  (solid black line). Panel (b): $(1 - F_{n})$ for different $I_n(t)$ and the same random sets of $\eps_{i>2}$ as in panel (a), where $F_n$ is the fidelity. The increased fidelity  as compared to the Grover Hamiltonian (horizontal black line)  is due to quantum interference. This effect is particularly dramatic for $n=3$ and $n=7$ with $1-F_3 = 5.20 \times 10^{-6}$ and $1-F_7 = 6.40 \times 10^{-5}$ for the random choice \#2 \cite{RandSeed}. The matrix size is $N = 64$ and the parameter $\delta = 0.1$ in the interpolating function $s(t)$. }
\label{Fig:Infid_Gaps_vs_n}
\end{figure}

As established in the previous section, when $\E$ is positive semi-definite with nonzero rank, all operators $-K_n$ share the same ground state $|\gamma\rangle$. Likewise, when $\eps_1 = 0$ and all other $\eps_i > 0$, the operators $\E^n$ share the common ground state $|1\rangle = |m\rangle$. It follows that the same adiabatic quantum computation can be implemented using any of the commuting $I_n(t)$ as the interpolating Hamiltonian,
\beg
I_n(t) = s(t)\, \E^n + \left[1 - s(t)\right] (-K_n).
\en
This raises a natural question: how does the choice of $I_n(t)$ affect the accuracy of the computation? To address this, we   compare the fidelities
\beg
F_n = \left| \left\langle m \middle| \Psi_{(n)}(T_\mathrm{run}) \right\rangle \right|^2,
\en
by numerically evaluating $\Psi_{(n)}(t)$, the solution of the non-stationary Schr\"odinger equation $i\, \partial_t \Psi = I_n(t)\, \Psi$, with the initial condition $|\Psi(0)\rangle=|\gamma\rangle$. For simplicity, we use the same interpolating function $s(t)$ given by \eref{Parametric_Interpol_LocalAdiabatic} for all $I_n(t)$.

    With the choice of $\eps_i$ in \eref{Quant_Search_Param_Choice}, all operators $I_n$, including $I_1 \equiv H = H_G$, are block-diagonal, consisting of a common $2 \times 2$ block and an $(N - 2) \times (N - 2)$ block proportional to the identity matrix. This $2 \times 2$ block governs the time evolution of the initial uniform superposition state $|\gamma\rangle$ under $I_n$, and hence the fidelities $F_n$ are identical for all $n$. To see this, note that
\beg
|\gamma\rangle = \frac{1}{\sqrt{N}} |1\rangle + \frac{\sqrt{N - 1}}{\sqrt{N}}\, |\tilde{2}\rangle,
\en
where $|\tilde{2}\rangle$ is a normalized state  orthogonal to $|1\rangle = |m\rangle$. The Hamiltonian $H_G$ in \eref{Ham_RC_LocalAdiabatic} is the identity matrix plus outer products of the vectors $|1\rangle$ and $|\tilde{2}\rangle$ with themselves and with each other. In the basis where $|1\rangle$ and $|\tilde{2}\rangle$ are the first two basis vectors, $H_G$ assumes the claimed block-diagonal form. \eref{in} implies that all $I_n$ inherit this structure, since raising $H + x\,\iden$ to any integer power preserves it.  

According to Eqs.~\re{eta} and \re{lambda}, the eigenvalues $\eta^{(n)}_{1}$ and $\eta^{(n)}_{2}$ of the $2 \times 2$ block of $I_n$, for the choice of $\eps_i$ in \eref{Quant_Search_Param_Choice} and $\gamma_i = 1/\sqrt{N}$, are given by
\beg
\eta^{(n)}_{1,2} = x \frac{N - 1}{N} \cdot \frac{1}{\lambda_{1,2} - 1},
\en
where $\lambda_1$ and $\lambda_2$ are the roots of the quadratic equation
\beg
\frac{1}{N} \left( \frac{1}{\lambda} + \frac{N - 1}{\lambda - 1} \right) = \frac{1}{x}.
\en
Since the $I_n$ commute, their $2 \times 2$ blocks must commute as well. Commuting $2 \times 2$ matrices with identical eigenvalues must coincide.

\begin{figure}
    \centering
    \includegraphics[trim={0.75cm 0.0cm 0.7cm 0.0cm},clip,scale=0.75]{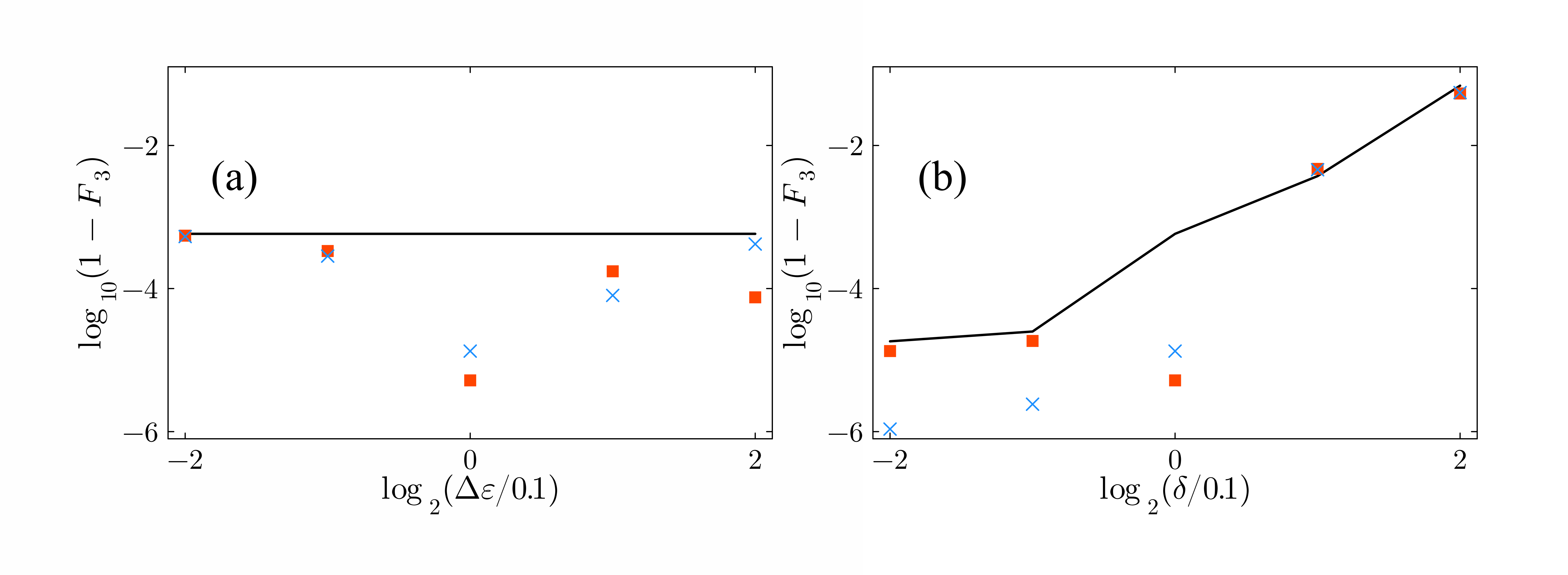}
    \caption{ Efficiency gain using $I_3$ for matrix size $N = 64$. Panel (a): Five values of $\Delta\eps$, spaced logarithmically, are tested with fixed $\delta = 0.1$. Random choices $\#1$ and $\#2$ are shown as blue crosses and red squares, respectively \cite{RandSeed}. As $\Delta\eps \to 0$, fidelity-deficit converges to that of the Grover Hamiltonian (black line). Notably, the performance of $I_3$ improves markedly at $\Delta\eps = 0.1$. Panel (b): Holding $\Delta\eps = 0.1$ fixed, we vary $\delta$ using the same random $\eps_i$ as in Fig.\ \ref{Fig:Infid_Gaps_vs_n}. As expected, both $T_\mathrm{run}$ and $F_3$ rise with decreasing $\delta$. Curiously, lowering $\delta$ brings out a marked efficiency gain in random choice $\#1$, driven by quantum integrability. }
\label{Fig:fid3_details}
\end{figure}

 We now turn to other choices of positive semi-definite $\E$ with $\eps_1 = 0$ and all other $\eps_i > 0$. As shown above, the same quantum search protocol can be implemented for any such $\E$ using any member $I_n$ of the commuting Type-1 family. Consider, for example, the following parameter choice:
\beg
\begin{split}
&\eps_1 = 0, \quad \eps_2 = 1, \\
&\eps_i = \text{uniformly distributed random numbers in the interval }  [1.0, 1.0 + \Delta\eps], \quad i > 2,
\end{split}
\label{Rand_eps_choice}
\en
where $\Delta\eps$ is a small positive number.   By construction, the energy gap between the ground and first excited states at $t = T_\mathrm{run}$ is the same (equal to 1) for all $I_n(t)$ and for the Grover Hamiltonian~\re{Ham_RC_LocalAdiabatic}. In addition, this gap is also fixed at 1 for all $I_1(t) \equiv H(t)$ at $t=0$, regardless of the specific choice of $\eps_i$ for $i > 2$. In this case, the Hamiltonian and the operators $I_n$ are no longer block-diagonal. Consequently, the instantaneous ground state is coupled not only to the first excited state but also to higher excited states, and the fidelities $F_n$ become distinct. As shown in Fig.~\ref{Fig:Infid_Gaps_vs_n} panel (a), the gap between the instantaneous ground state and the first excited state is nearly the same as that for the 
Grover Hamiltonian~\re{Ham_RC_LocalAdiabatic} across all $I_n$.

At first glance, one might expect that quantum tunneling to higher excited states would reduce the fidelity by increasing probability leakage out of the ground state. Surprisingly, our numerical simulations indicate the opposite behavior; see Fig.~\ref{Fig:Infid_Gaps_vs_n}. The fidelity $F_n$ exceeds that of the Grover Hamiltonian~\re{Ham_RC_LocalAdiabatic} for all $I_n$, and for $n = 3$, fidelity-deficit $1 - F_n$ is smaller by {\em two orders of magnitude}. This enhancement is a result of quantum interference: probability amplitudes for transitions out of the ground state combine destructively, suppressing leakage into excited states.

Fig.~\ref{Fig:fid3_details} examines $F_3$ more closely for $n = 3$, $N = 64$. In panel (a), two sets of random $\eps_{i>2}$ are drawn for various $\Delta\eps$, keeping $\delta = 0.1$ fixed. As $\Delta\eps \to 0$, the fidelity approaches that of the Grover Hamiltonian. In this instance, the best performance occurs for random choice $\#2$ at $\Delta\eps = 0.1$. Panel (b) explores the effect of varying $\delta$ at fixed $\Delta\eps = 0.1$. As expected, lowering $\delta$ improves fidelity --- at the cost of a longer $T_\mathrm{run}$.

Recall that the functional form of the interpolating schedule depends on two factors: (1) the energy gap $\left(\eta_{2} - \eta_{1} \right)$ between the instantaneous ground state $\Psi_0(t)$ and the first excited state $\Psi_1(t)$ and (2) the transition matrix element $\langle \Psi_1(t) | d I_n(s)/ ds | \Psi_0(t)\rangle = \langle \Psi_1(t) | \left(\mathcal{E}^{n} + K_n\right)  | \Psi_0(t)\rangle$ \cite{LAA}. In particular, the anneal rate bound reads 
\beg
\begin{split}
\left|\frac{ds(t)}{dt}\right| \leqslant \delta \frac{\left(\eta_{2} - \eta_{1} \right)^2}{\left|\left\langle \Psi_1(t) \left| \frac{d I_n(s)}{ds} \right| \Psi_0(t)\right\rangle\right|}.
\end{split}
\label{Anneal_Speed}
\en 
Using the Cauchy-Schwartz inequality, one obtains that $\langle \Psi_1(t) | d I_n(s)/ ds | \Psi_0(t)\rangle = \mathcal{O}(1)$. Following the Grover case \cite{LAA}, we set $\left|\left\langle \Psi_1(t) \left| d I_n(s)/ ds \right| \Psi_0(t)\right\rangle\right| = 1$ to (effectively) saturate the bound in Eq.\ \eqref{Anneal_Speed}. Our numerical results confirm the assertion a posteriori. Had the transition matrix element’s magnitude substantially exceeded unity, achieving comparable fidelities $F_n$ would have required a much slower anneal rate. Since $(\eta_{2} - \eta_{1})$ is approximately the same as for the Grover Hamiltonian (see Fig.\ \ref{Fig:Infid_Gaps_vs_n}), we use a single interpolating function for all $I_{n}(t)$. It follows that $T_\mathrm{run} \propto \sqrt{N}$ for every interpolating Hamiltonian, and the higher Type-1 matrices deliver a constant-factor improvement over the Grover Hamiltonian. 

It is worth noting that Type-1 matrices also give rise to certain integrable multi-level Landau-Zener models~\cite{Patra_2015,Yuzbashyan_2018}. These models are known to exhibit similarly favorable quantum interference effects among transition amplitudes~\cite{demkov,Sinitsyn_2015,Sinitsyn_2017}. These results indicate that exploiting the integrability of Type-1 matrices within adiabatic quantum computation can meaningfully enhance performance. The commuting hierarchy supplies a systematic route to new interpolating Hamiltonians, furnishing an additional knob for tuning Grover-style search efficiency. It is also natural to ask whether controlled departures from integrability might yield further gains. 

To conclude this section, we highlight several features relevant for physically implementing the analog Grover search using Type-1 commuting partners. Realizing the Grover Hamiltonian for a database of size $N = 2^{k}$ with $k$ qubits inevitably generates up to $k$-body interactions that are long-range under any geometry \cite{QAA_Theory}. These include terms of the form $\prod_{i \in \mathbb{S}} \hat{\sigma}_{z_{i}}$ and $\prod_{i \in \mathbb{S}} \hat{\sigma}_{x_{i}}$, where $\hat{\sigma_z}$ and $\hat{\sigma_x}$ are Pauli operators and $\mathbb{S} \subseteq \{1, 2, \ldots, k\}.$ One partial route toward physical realizability is provided by the so-called ``gadget Hamiltonians'', which replace such high-order interactions by enlarged Hamiltonians with only simple (typically two-body) couplings, while reproducing the target Hamiltonian within a low-energy sector. Although gadgetized constructions \cite{Gadget_1, Gadget_2, Gadget_3, Gadget_4, Gadget_5, Gadget_6} and several other proposed physical implementations of the Grover Hamiltonian \cite{Grover_Expt_Implement_1, Grover_Expt_Implement_2} involve only two-body interactions, the resulting Hamiltonians remain intrinsically long-range. These Hamiltonians are structurally analogous to Gaudin magnets, where a distinguished degree of freedom couples collectively to the rest of the system \cite{Gaudin}. In this light, we note that the commuting partners of the Type-1 family --- specifically the operators $Z_{j}$ in Eq.\ \eqref{Z-s} --- are themselves equivalent to Gaudin-type Hamiltonians \cite{haile}. This observation clarifies that extending the analog Grover protocol using higher integrals $I_{n}$ does not introduce qualitatively new implementation challenges beyond those already present for the original Grover Hamiltonian. We therefore expect the mapping presented in Ref.\ \cite{haile} along with Eq.~\eqref{Im-as-Z} to provide a practical route for realizing the full Type-1 commuting hierarchy with comparable experimental overhead.

\section{Concluding Remarks} 
\label{Conclusions}

 In this work, we have noted that Weyl's relations involving two unitary matrices $A$ and $B$ in $N$ dimensions can be usefully generalized to include a third matrix $C$. The matrix $C$ has interesting commutation relations with $\log A$ which permits one to view it as a finite dimensional version of the operator $\hat{Q} \hat{P}+ \hat{P} \hat{Q}$ in quantum mechanics, while its commutation relation with $B$ enables us to view it as the canonical momentum $\hat{P}$ when restricted to an $(N-1)$-dimensional subspace obtained by eliminating a specific state of the original state space. A simple procedure for modifying arbitrary operators (matrices) to annihilate the specific state then generates an effectively finite-dimensional canonical theory.   These matrices are then used to construct  a hierarchy of mutually commuting operators, depending on   $\sim 2 N$ independent real parameters. These operators are shown to be intimately connected to  Type-1 matrices developed by us earlier \cite{yuzbashyan}, using very different and independently developed  ideas. The Type-1 matrices provide  a realization of quantum integrable systems --- such as the Heisenberg antiferromagnetic spin chain, or the one-dimensional Hubbard model, so that the new hierarchy can be viewed in the same light.

 Additionally, we find that the hierarchy of commuting operators obtained here contain the leading member $I_1$, given by \disp{our-H}, which, for a certain choice of the parameters, is identical to the Grover Hamiltonian~\re{Ham_RC_LocalAdiabatic}, widely studied in the field of quantum computation. We present  results that suggest a significant advantage of employing higher members $I_n$ with $n\geq2$ in similar studies, which seem to be worth pursuing further.

 Finally, we find it most remarkable  that Weyl's ideas \cite{Weyl} continue to nourish current research and open new directions,   nearly a century after their original publication.

\begin{acknowledgments}
We are grateful to Dominik \v{S}afr\'{a}nek for bringing Ref.\ \cite{LAA} to our attention. AP acknowledges the financial support from the Institute for Basic Science (IBS) in the Republic of Korea through Project No. IBS-R024-D1. 
\end{acknowledgments}

\appendix 
\section{Deriving the Heisenberg Algebra from the Weyl Matrices by Taking the $N\to \infty$ Limit}
\label{Hesenberg_Derivation_Details}

In this Appendix, we start with the pair of  $N\times N$  unitary matrices $A$ and $B$ as introduced in Section \ref{N-finite}. We then lay out the derivation of the Heisenberg algebra in full. 

Following Schwinger \cite{Schwinger2}, we make a symmetric choice,  
  \beq
  \xi= \eta=\sqrt{\omega_0} \label{Weyl-3}.
 \eeq
 We must also deal with the situation that the eigenvalues of $\log B$, namely $i \omega_0 n$, are discrete with a small spacing, while the eigenvalues of $\hat{Q}\to x$ are quasi-continuous --- indeed, we would like to take the continuum limit as $N\to \infty$. This is achieved by the Weyl scaling, where
 \beq
 x\leftrightarrow \sqrt{\omega_0} \; n, \quad \textrm{and} \quad |x\rangle  \leftrightarrow \frac{1}{\sqrt[4]{\omega_0} } |n\rangle,  \label{Weyl-5} 
 \eeq 
 so that lattice constant  $\sqrt{\omega_0}$   shrinks as $N$ increases. With \beg
 x_1=\sqrt{\omega_0}\, n, \quad x_2=\sqrt{\omega_0}\, m,
 \label{x1x2}
 \en
 the Dirac delta overlap of the continuum states is related to the Kronecker delta overlap in the discrete case through the relation
 \beq
 \langle x_1|x_2\rangle= \delta(x_1-x_2)\leftrightarrow  \frac{1}{\sqrt{\omega_0}} \delta_{n,m}.
 \label{Weyl-4}
 \eeq
  
A completely parallel situation exists for the momentum operator $\hat{P}$, which also has a continuous set of eigenvalues in the large $N$ limit case. We skip its discussion since we only need \disp{Weyl-3} and \disp{Weyl-4} below.

On the other hand,  we now regard $N$ as finite and compute the logarithms of the  matrices and evaluate their commutators. We begin with
\beq
\log B= i  \omega_0 \sum_{n=0}^{N-1} n \; |n\rangle \langle n|   \label{logB}
\eeq  
and
\beq
\log A= i \omega_0 \sum_{r=0}^{N-1} r |k_r\rangle \langle k_r| \label{logA-1}
\eeq
Using \disp{k-state},  we obtain
\beq
\log A&=& i \pi \left(1- \frac{1}{N}\right) \iden + i \frac{\pi }{N}\sum_{n\neq m}\frac{|n\rangle \langle m| b_m-|m\rangle \langle n| b_n }{b_n-b_m},  \label{logA-20} \\
&=& i \pi \left(1- \frac{1}{N}\right) \iden + i \frac{\pi }{N} \sum_{n \neq m} \left[ \cot \frac{\pi}{N}(n-m) -i \right] |n\rangle \langle m|   , \label{logA-2}
\eeq
the second form \eqref{logA-2} illustrating the useful result  that convolution with the cotangent in real space is equivalent to the first derivative for a class of functions on the lattice \cite{Shastry-inverse}.

Going further, we compute the matrix element of the commutator \cite{comment-Santhanam}. Denoting   $\Delta= \omega_0 (n-m)$, and treating $\Delta$ formally as a continuous variable,  we obtain the matrix element
\beq
\langle n|\, [\log B,\log A]\, |m \rangle&=&- \frac{\omega_0}{N}   {\Delta} \sum_{r=0}^{N-1} r e^{ i \Delta r} \nn \\
&=& i \frac{\omega_0}{N}   {\Delta} \frac{d}{d \Delta} \left( e^{ i \half \Delta (N-1)}  \frac{\sin  \half \Delta N}{\sin \half \Delta}\right). \label{Commute-discrete}
\eeq
Further,  \eref{x1x2} implies $\Delta=\sqrt{\omega_0} (x_1-x_2)$  and  using \disp{Weyl-3} we rewrite \disp{Commute-discrete} as
\beq
\sqrt{\omega_0} \langle x_1| [\log B, \log A]  | x_2 \rangle=  i \frac{\omega_0}{N}   {(x_1-x_2)} \frac{d}{d (x_1-x_2)} \left( e^{ i c_0 (x_1-x_2) } \frac{\sin  \sqrt{\half \pi N} (x_1-x_2)}{\sin  \sqrt{\frac{ \pi}{2 N}} (x_1-x_2) }\label{first} \right)\!\!, \label{comm-3}
\eeq
where $c_0=\half \sqrt{\omega_0}  (N-1)$. By using  Dirichlet-Bonnet's result \cite{Dirichlet-Bonnet}, we can set 
\beq
\lim_{N\to \infty} \frac{\sin  \sqrt{\half \pi N} (x_1-x_2)}{\sin  \sqrt{\frac{ \pi}{2 N}} (x_1-x_2) }\to \frac{2 \pi}{\sqrt{\omega_0}} \delta(x_1-x_2).
\eeq
Upon using $(x_1-x_2) \delta'(x_1-x_2)=-\delta(x_1-x_2)$, \disp{comm-3}  reduces to
\beq
\sqrt{\omega_0} \langle x_1| [\log B, \log A]| x_2 \rangle=  - i (\omega_0)^{\frac{3}{2}}   \delta(x_1-x_2).
\eeq
Using the replacements $\log B\to i \sqrt{\omega_0} \hat{Q}$ and $\log A \to i  \sqrt{\omega_0} \hat{P}/\hbar$ and canceling the common $\omega_0^{\frac{3}{2}}$ term, we recover the matrix elements of the Heisenberg algebra~\re{Heisenberg},
 \beq
\langle x_1| [\hat{Q},\hat{P}] |x_2\rangle= i \hbar \delta(x_1-x_2). \label{Heisenberg-mat}
\eeq


\begin{thebibliography}{99}
\bibitem{gamma} Here we have anticipated the connection between the uniform superposition and the projection operator $\P$, which is defined via the state $|\gamma\rangle$ as $\P = |\gamma \rangle \langle \gamma|$.
\bibitem{RandSeed} In the programming language Julia, which we used to generate the figures, fixing $\Delta\eps$ and an integer seed determines a specific set of random values for $\eps_{i>2}$. We refer to seed values 1 and 2 as “Random Choice \#1” and “Random Choice \#2,” respectively.
\bibitem{comment-Santhanam} Weyl~\cite{Weyl} notes on p. 255: 
``The  commutation rules which are satisfied by $q,p$ in quantum mechanics cause some trouble.''  He outlines a general procedure for taking the continuum limit to obtain the Heisenberg commutation relations  ``symbolically'' written as $q(x) p(x')-p(x') q(x)= i \delta(x-x')$, see Eq. (12.3) in \cite{Weyl}, with the notation (p. 254): $\delta(x-x')= \frac{1}{\Delta x} \delta_{x x'}$, and $\Delta x=\frac{1}{N}$, with $\Delta x \to 0$ in the limit of large $N$. In the present case  it is appropriate to take $\Delta x=\frac{\sqrt{2 \pi}}{\sqrt{N}}$,  leading to  \disp{Weyl-5} and \disp{Weyl-4}.  With this choice of $\Delta x$, also vanishing  in the large $N$ limit, the matrix element equation~\re{Heisenberg-mat} leads systematically  to the Heisenberg commutator~\re{Heisenberg}, as shown. We term this refined procedure as Weyl's scaling.
\end{thebibliography}
\end{document}